\documentclass[11pt]{amsart}
\usepackage{amsfonts,amsmath,eucal}
\usepackage[all]{xy}
\begin{document}

\newcommand{\A}{{\mathbb A}}
\newcommand{\bA}{{\overline{\mathbb A}}}
\newcommand{\C}{{\mathbb C}}
\newcommand{\LL}{{\mathbb L}}
\newcommand{\PP}{{\mathbb P}}
\newcommand{\Q}{{\mathbb Q}}
\newcommand{\R}{{\mathbb R}}
\newcommand{\Z}{{\mathbb Z}}
\newcommand{\FO}{{{\rm FO}^3}}
\newcommand{\M}{{\overline{\mathcal M}}}
\newcommand{\la}{{\langle}}
\newcommand{\ra}{{\rangle}}
\newcommand{\Config}{{\rm Config}}
\newcommand{\End}{{\rm End}}
\newcommand{\Hom}{{\rm Hom}}
\newcommand{\HH}{{\rm HH}}
\newcommand{\tLambda}{{\widetilde{\Lambda}}}
\newcommand{\PSl}{{\rm PSl}}
\newcommand{\pt}{{\rm pt}}
\newcommand{\MM}{{\bf M}}
\newcommand{\XX}{{\bf X}}

\title{The Bagger-Lambert model and Type IIA string theory}
\author{Jack Morava}
\address{The Johns Hopkins University,
Baltimore, Maryland 21218}
\email{jack@math.jhu.edu}

\subjclass{18D50, 53D35, 53D37}
\date{15 May 2014}
\begin{abstract}{Notes from a talk at the Workshop on Geometry, Topology, and Physics 
at the University of Pittsburgh, 14-15 May 2014
\[
{\tt http://www.mathematics.pitt.edu/node/1216}
\]
We conjecture the existence of a `compactified' version of Fukaya's homology for
symplectic manifolds, which carries a canonical 2-Gerstenhaber algebra structure.
This may help to understand the 2-Lie algebra structure involved in models [2] for
interacting D-branes.}\end{abstract}
\bigskip

\thanks{Thanks to Jon Bagger, Satyan Devadoss, Nitu Kitchloo, and Jim Stasheff for help
and advice!}
 
\maketitle \bigskip

\section{Introduction} \bigskip

\noindent
This {\bf tentative} sketch was inspired by a very interesting recent review article [2] about
multiple $D$-branes in $M$-theory. There I learned, for example, that:

$\; \bullet \;$ the {\bf low}-energy (`supergravity') limit of $M$-theory involves {\bf geometric} 
objects: ambient space-time, branes, \dots; but  

$\; \bullet \;$ $M$-theory is thought to have {\bf no} coupling constants, and hence has no natural 
candidate for a perturbative approximation.

$\; \bullet \;$ Nevertheless, it is thought to specialize to the classical string theories (type I, IIA, 
IIB,\dots) in the right circumstances. \bigskip

\noindent
The Bagger-Lambert model (which is currently apparently not particularly fashionable) aspires to accomodate 
{\bf wrapping} of branes and related phenomena, which may lack any very familiar interpretation in geometric 
topology: branes, as they appear in $M$-theory, are analytic objects, which makes intuitions 
about issues such as transversality problematic. \bigskip

\noindent
The supergravity limit of the BL model is formulated in terms of Lie 2-algebras [HW], involving a 
triple bracket $\la -,-,- \ra$ satisfying an analog of the Jacobi identity. Together with a more classical 
graded-commutative algebra, this defines a 2-analog of a Gerstenhaber algebra.

\newpage

\noindent
The Bagger-Lambert-Gustavsson action is \bigskip
 
\[
{\bf something \; like \; an \; 11D \; supergravity \; Lagrangian}
\]
\[
+
\]
\[
{\bf interaction \; terms \; like} \; |\la X,X,X \ra|^2 \;.
\] \bigskip

\noindent
In interesting examples the Lie 2-algebra structures can be reinterpreted in terms of classical Lie 
algebras, resulting in more familiaar-looking models involving Chern-Simons-type lagrangians for 
(perhaps unexpected) combinations of gauge groups.\bigskip

\noindent
The realization that many classical 2D conformal field theories (eg associated to free loopspaces of
manifolds) have Gerstenhaber algebra structures arising from the homology of an action of the topologists'
little 2-disks operad [12 \S 7.4] profoundly affected the later developments in such theories [14 \S 2.4]. 
\bigskip

\noindent
{\bf The aim of this talk} is to show how certain generalized Type IIA (ie symplectic Fukaya) models 
manifest natural Gerstenhaber 2-algebra structures (arising from certain underlying operad actions), 
and to suggest that this may reflect an action of some kind of homotopy Gerstenhaber 2-algebra structure 
on the underlying algebras of differential forms on the space-time background. \bigskip

\section{Gromov-Witten invariants in Type IIB string theory and (small) quantum 
cohomology}\bigskip

\noindent
{\bf 2.1} The Deligne-Knudsen-Mumford moduli stack $\M_{g,n}(\C)$ of genus $g$ stable (ie with 0-dimensional 
Lie algebra of automorphisms) complex algebraic curves, marked with $n$ distinct smooth points ($n \geq 3$ 
if $g=0$, $\geq 1$ if $g=1$), has a canonical stratification indexed by certain abstract weighted (`modular') 
connected graphs (with $n$ external vertices). When $g=0$ these are (abstract, unrooted) trees.\bigskip

\noindent
Gluing two such curves together at chosen marked points defines morphisms
\[
\M_{g,n+1} \times \M_{h,1+m} \; \to \; \M_{g+h,n+m}
\]
which make the collection $\{\M_{g,n}\}$ into a generalized (`modular') operad; but this sketch will be 
concerned with the classical operad defined by the subcollection $\{\M_{0,n+1}\} \; (n \geq 2)$ and its 
suitably associative maps
\[
\M_{0,n+1} \times \prod_{1 \leq k \leq n} \M_{0,i_k+1} \; \to \; \M_{0,\sum i_k +1}
\]
(corresponding to the grafting of rooted trees). These spaces have torsion-free homology; moreover

1) forgetting to distinguish a marked point makes both of these (generalized) operads into {\bf cyclic} 
operads [10], and 

2) without new ideas about $\M_{0,2}$, these operads are non-unital [16].\bigskip

\noindent
{\bf 2.2} Now suppose that $V$ is a complex projective smooth algebraic variety, `convex' in a
certain sense; then there are [BM \dots] moduli stacks (or orbifolds) $\M_{g,n}(V)$ of stable 
(ie with 0-dimensional Lie algebras of automorphisms [13]) curves
\[
\phi : C_{g,n} \to V
\]
in $V$, together with gluing maps generalizing the case $V = \pt$ above, as well as 
{\bf evaluation} morphisms
\[
GW : \M_{g,n}(V) \; \to \;  \M_{g,n} \times V^n \;.
\]
The Gromov-Witten invariant 
\[
GW_{g,n}(V)  \in H_*(\M_{g,n} \times V^n,\Lambda)
\]
is the cohomology class (with coefficients in the rational group ring $\Lambda = \Q[H_2(V,\Z)]$, 
perhaps completed) defined by this (locally algebraic) cycle, with its components weighted by their 
degrees $d(\phi) \in H_2(V,\Z)$.\bigskip

\noindent
{\bf 2.3} It will often be useful below to interpret a map $A \to X \times Y$ as a 
geometric correspondence $\xymatrix{A: X \ar@{.>}[r] & Y}$. In the category of smooth compact 
oriented manifolds and maps, Poincar\'e duality defines an associated  homomorphism 
\[
[A] : H^*(X) \; \to \; H^{*+d}(Y), \; d := \dim Y - \dim A
\]
of cohomology groups (with coefficients in, say, a ring allowing a K\"unneth isomorphism);
details of such constructions are summarized below in an appendix.
Duality then allows us to interpret Gromov-Witten invariants as elements $GW_{g,n+1}(V)$ in
\[
\Hom^*_\Lambda(H_*(\M_{g,n+1}),\Hom(H^*(V)^{\otimes n}, H^*(V)))
\]
(where cohomology, from now on, has coefficients in the Novikov ring $\Lambda$). \bigskip

\noindent
In a Cartesian closed category
\[
\End_n(X) := {\rm Maps}(X^n,X)
\]
defines the endomorphism operad of $X$, and the associativity properties of pointwise gluing imply 
that
\[
GW_{0,*} : H_*(\M_{0,\bullet+1},\Lambda) \to \End_\bullet^\Lambda(H^*(V,\Lambda))
\]
is a morphism of operads: thus $H^*(V,\Lambda)$  becomes an algebra over the 
(small) quantum cohomology operad $\{H_*(\M_{0,\bullet+1},\Lambda)\}$.  \bigskip 

\noindent
{\bf 2.4} As an application, Manin's polycommutative product [9]
\[
v *_t w :=  \sum_{n \geq 1} [\M_{0,n+2}](v \otimes w \otimes t \otimes \cdots \otimes  t)
\]
($t,v,w \in H^*(V,\Lambda)$, with $t$ repeated $n$ times in the product on the right) makes the
cohomology of $V$ into a Frobenius manifold. Taking higher genus terms into account defines 
something like a Frobenius manifold structure on the complex cobordism of $V$ [17]. \bigskip

\section{Type IIA strings, Fukaya's category, and Devadoss's mosaic operad}\bigskip

\noindent
At this point we start over, now with $(V,\omega)$ a compact symplectic manifold: \bigskip

\noindent
{\bf 3.1 Definition} A {\bf Lagrangian polygon} (cf [$\FO$])
\[
\LL \; :=  \; \la L_1, \dots, L_n \ra
\]
in $V$ consists of 

$\; \bullet \;$ oriented Lagrangian submanifolds $L_1, \dots, L_n$ of $V$ (cyclically ordered for 
convenience), such that $L_i$ intersects $L_{i+1}$ transversally, 

$\; \bullet \;$ a pseudoholomorphic map $F : D \to V$ from the closed two-disk to $V$, together with a 
choice $\{z_1,\dots,z_n\} \subset \partial D = \PP_1(\R)$ of $n$ distinct points on the boundary of the 
disk, such that $F(z_i) \in L_i \cap L_{i+1}$, 

$\; \bullet \;$ such that $F$ maps the interval $I_k := [z_i,z_{i+1}] \subset \partial D$ to $L_i$, 

$\; \bullet \;$ satisfying a relative spin condition:
\[
w_2(T_{L_i}) \in {\rm image} \; [H^2(V,\Z_2) \to H^2(L_i,\Z_2)] \;.
\]

\noindent
A morphism $\LL \to \LL'$ of Lagrangian polygons is a commutative diagram
\[
\xymatrix{ 
D \ar[d]^F \ar[r]^\phi & D' \ar[d]^{F'} \\
V \ar[r]^\Phi & V }
\]
in which $\Phi$ preserves $\omega$, while $\phi$ is holomorphic, taking the boundary decomposition
of $\LL$ to that of $\LL'$. There is then a topological stack $\A_n(V)$ of such Lagrangian $n$-gons 
in $V$, with invertible maps of such polygons as morphisms. \bigskip

\noindent
{\bf 3.2 Remarks:} \bigskip

\noindent
1) There is an implicit action of the dihedral group of order $2n$ on $\A_n$: the cyclic group of order
$n$ acts by shifting the labels on the $L_k$'s, while reversing their order takes the category $\A_n$ into
itself, perhaps reversing its orientation.\bigskip

\noindent
2) Fukaya defines an algebra structure on the free $\Z$-module generated by 
equivalence classes $[L]$ of oriented Lagrangian submanifolds of $V$, with product
\[
[L_i] \cdot [L_j] = \sum \Psi^k_{ij} [L_k] \;,
\]
where the coefficients $\Psi^k_{i,j} := \# \la L_i,L_j,L_k \ra$ count the number of Lagrangian triangles 
bounded by the indicated Lagrangians -- under the expectation that the space of such things is a zero-dimensional 
oriented manifold. \bigskip

\noindent
3) Note that we can integrate $\omega$ over $\LL$ to obtain its area $\omega(\LL)$.\bigskip

\noindent
4) It will simplify notation below to write $\{I_*\}$ for the ordered partition $I_1, \dots,I_k$ of the projective
line into intervals.\bigskip

\noindent
{\bf 3.3 Definition} The locus $\M_{0,n}(\R) \subset \M_{0,n}(\C)$ of real points on the moduli stack 
of genus zero curves marked with $n$ distinct smooth points can be identified with a compactification 
\[
\Config^n(\PP_1(\R))//\PSl_2(\R) := \M_{0,n}(\R)
\]
of the quotient of the space of distinct $n$-tuples on the real projective line, under projective 
equivalence. Its elements can be regarded as (possibly decomposed) hyperbolic $n$-gons in the Poincar\'e 
disk, with geodesic boundaries, having all vertices on $\PP_1(\R)$ and one in particular at $\infty$. 
The collection $\{\M_{0,\bullet+1}(\R)\}$ defines Devadoss's {\bf mosaic} operad [5]; relaxing the choice 
of vertex at infinity makes it a cyclic operad. \bigskip 

\noindent
{\bf 3.4.1 Conjecture}, cf [18, 19]: Under reasonable hypotheses on $V$, there are completions
\[
\A_n(V) \subset \bA_n(V)
\]
constructed by adjoining strata of decomposed Lagrangian polygons, indexed by {\bf planar}
trees, together with maps of these polygons to $V$ which are holomorphic on the interiors of its 
components, and continuous on their boundaries. Evaluation defines (Fredholm) maps
\[
\bA_n(V) \to \M_{0,n}(\R) \times V^{\{I_*\}}
\]
of topological groupoids; where 
\[
V^{\{I_*\}} = \prod_{1 \leq k \leq n} V^{I_k} 
\]
(note that the space $V^{I_k}$ of free maps of the interval $I_k$ to $V$ is homotopy equivalent to $V$ itself). 
\bigskip

\noindent
{\bf 3.4.2} Moreover, these maps satisfy an associativity condition, which requires some abbreviation to display:
\[
\xymatrix{
\bA_{n+1} \times_{X^{\{I(i_*+1)_1\}}} \prod \bA_{i_*+1} \ar[d] \ar[r] & \bA_{\sum i_k +1} \ar[d] \\
\MM \times_{\prod X^{I(i_*+1)_1}} \XX \ar[r] & \M_{0,\sum i_k+1}(\R) \times \XX' }
\]
where
\[
\MM := \M_{0,n+1}(\R) \times \prod \M_{0,i_k+1}(\R),
\]
\[
\XX := \prod X^{\{I(i_k+1)_*\}} \times X^{I(n+1)_{n+1}},
\]
and 
\[
\XX' := \prod X^{\{I(i_k)_*\}} \times X^{I(n+1)_{n+1}} \;.
\]

\noindent
Roughly speaking, then, we have geometric correspondences
\[
\xymatrix{
\bA_{\bullet +1} : \M_{0,\bullet +1}(\R) \ar@{.>}[r] & \End_\bullet(X^I) }
\]
which define an $\{H_*(\M_{0,\bullet +1}(\R),\tLambda)\}$-algebra structure on $H^*(V,\tLambda)$:
where now 
\[
\tLambda := \Q\{\{q\}\}[t]
\]
is an algebra over a field of Puiseux series in $q = \exp(\hbar)$, with components of $\bA_{n+1}$ 
weighted by $\exp(\omega(\LL)\hbar)t^n$.\bigskip

\section{Hochschild homology of $A_\infty$ ringspectra}\bigskip

\noindent
{\bf 4.1} The fundamental geometric fact about the moduli spaces $\{\M_{0,\bullet}(\R)\}$ is 
that they are {\bf aspherical}. They are tesselated
\[
\xymatrix{
\Sigma_{n+1} \times_{D_{n+1}} K_n \ar@{>>}[r] & \M_{0,n+1}(\R) }
\]
by Stasheff associahedra $K_n$, defining a piecewise negatively curved metric which implies them
to be spaces of type $K(\pi,1)$; thus, for example, $\M_{0,5}(\R)$ is Kepler's Great Dodecahedron. 
\bigskip

\noindent
Work of Etinghof, Henriques, Kamnitzer and Rains [6 Theorem 2.14] shows that algebras over the 
(unital) operad $\{H_*(\M_{0,\bullet + 1}(\R),\Q)$ are rational 2-Gerstenhaber algebras. On the
other hand,
\[
H_*(\M_{0,\bullet + 1}(\R),\Z_2) \; \cong \; H_{*/2}(\M_{0,\bullet +1}(\C),\Z_2) \;.
\]
When $n > 3$ Devadoss's spaces are non-orientable, and the homology of their orientation covers is
not yet understood. \bigskip

\noindent
{\bf 4.2} The action of the dihedral group $D_{n+1}$ appeared in \S 3.2 above. By regarding $\Sigma_{n+1}$ 
as $\Sigma_n \cdot C_{n+1}$, the presentation above defines the structure
\[
\Sigma_n \times K_n \to (\Sigma_n \cdot C_{n+1}) \times_{D_{n+1}} K_n \to \M_{0,n+1}(\R)
\]
of an $A_\infty$ space on the collection $\{\M_{0,\bullet + 1}(\R)\}$, permitting us to interpret
$H^*(V,\tLambda)$ as an $A_\infty$ algebra (but unitality (cf \S 2.1.2, [15 \S 5.5.7]) deserves further 
attention \dots). \bigskip

\noindent
Angeltveit [1] has defined a generalized Hochschild homology for $A_\infty$ ringspectra. I will close by
noting that the resulting
\[
\HH_*(H^*(V,\tLambda))
\]
seems related in interesting ways to the symplectic cohomology of $V$ defined recently by Ganatra [8].
The mod two analog of this construction, and possible variants defined using Devadoss's orientation covers
of $\M_{0,\bullet + 1}(\R)$, have yet to be considered. \bigskip

\section{Appendix on conventions}\bigskip 

\noindent
{\bf Correspondences} (re \S 2.3)

\[
\xymatrix{A : X \ar@{.>}[r] & Y} \Longleftrightarrow A \to X \times Y \;,
\]
\[
\xymatrix{B : Y \ar@{.>}[r] & Z} \Longleftrightarrow B \to Y \times Z \;,
\]
$\Rightarrow$
\[
\xymatrix{A \times_Y B \ar@{.>}[r] & X \times Z}
\]
defined by
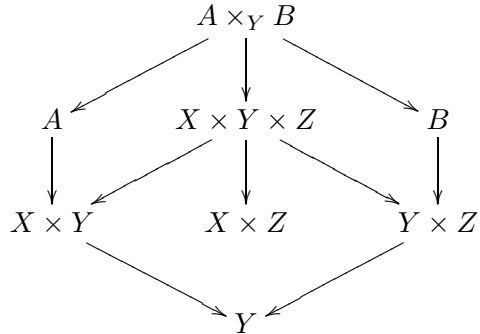
\[
\xymatrix{
{} & \ar[dl] A \times_Y B \ar[d] \ar[dr] & {} \\
A \ar[d] & \ar[dl] X \times Y \times Z \ar[d] \ar[dr] & B \ar[d] \\
X \times Y \ar[dr] & X \times Z  & \ar[dl] Y \times Z \\
{} & Y & {} } \;.
\]

\noindent
Note, the category of correspondences is {\bf self-dual}.\bigskip

\noindent
{\bf Example} Hecke correspondences are defined by morphisms
\[
G \to H \times K
\]
of (finite?) groups \dots \bigskip

\noindent
Note, if the objects involved are Poincar\'e-duality objects, then
a correspondence $\xymatrix{A : X \ar@{.>}[r] & Y}$ defines (assuming 
a K\"unneth formula)
\[
[A] \in H^*(X \times Y) \cong {\rm Hom}(H^*(X),H^*(Y))
\]
satisfying $[A \times_Y B] = [A] \circ [B]$. \bigskip

\noindent
{\bf Indexing conventions} (re \S 3.4): In fiber products of the form
\[
\bA_{n+1} \times_{X^I} \bA_{1+m}
\]
the parametrization of the boundary segment $[z_n,z_{n+1}]$ on the left is identified (via the action of the
projective group) with the parametrization of the segment $[z_1,z_2]$ on the right: thus the iterated fiber
product on the top left of the diagrams involves identifications over a product of the form
\[
\prod_{1 \leq k \leq n} X^{I(i_k+1)_1} \;,
\]
with $I(i_k+1)_1$ being the first interval in the partition $\{I(i_k+1)_*\}$ of $\PP_1(\R)$ defined by the
polygon $\LL_{i_k+1}$. In \S 3.4.2, the terms in the fiber products have been regrouped for readability, 
using a telescoping product identification of the form
\[
\pt \times_{\prod X^{I(i_*+1)_1}} \prod X^{\{I(i_*+1)\}} \cong \prod X^{\{I(i_*)\}} \;.
\]

\noindent
Note also that composition of correspondences involves maps of the form
\[
(X \times Y) \times_Y (Y \times Z) \cong X \times Y \times Z \to X \times Z
\]
which `cancel' paired copies of $Y$. After taking cohomology, these cancellations corresponds to 
applications of the trace map
\[
H^*(Y,k) \otimes H^*(Y,k) \to k \;.
\]
\bigskip

\bibliographystyle{amsplain}

\end{document}